\documentstyle[12pt,epsf,aps]{revtex}

\newcommand{\be}{\begin{equation}}
\newcommand{\ee}{\end{equation}}
\newcommand{\bea}{\begin{eqnarray}}
\newcommand{\eea}{\end{eqnarray}}
\begin{document}

%-----------------------------------------------------------

\title{
\begin{flushright}
{\normalsize hep-th/0008187 \\}
\end{flushright}
Effective Lagrangian for the pseudoscalars interacting 
with photons in the presence of a background electromagnetic field
\vspace*{7mm}
}
\author{K. V. Shajesh\footnote{Email: kvshajesh@yahoo.com}}
\address{
Physical Research Laboratory,
Navrangpura, Ahmedabad - 380 009, India.
}

\maketitle
\vspace{2cm}

%-----------------------------------------------------------

\begin{abstract}
We attempt to evaluate the effective
Lagrangian for a classical background field interacting with the
vacuum of two quantum fields.
The integration of one of the quantum fields
in general leads to a non-local term in the effective 
Lagrangian and thus becomes intractable
during the integration of the other quantum field.
We show that $\phi F \tilde{K}$ interaction is an exception.
We present the complete calculation for the evaluation of the 
effective Lagrangian for a pseudoscalar field interacting 
with photons in the presence of a background electromagnetic field.
Expression for the probability of the vacuum breaking down 
into a pseudoscalar-photon pair is evaluated.
We conclude that the gradient of an electric field 
beyond a certain threshold can give
rise to pseudoscalar-photon pair production.
\end{abstract}

%------------------------------------------------------------
\newpage
%\begin{multicols}{2}

\section{Introduction}
{\label{intro}}

Consider an interaction involving three fields.
Let us approximate one of the fields to be a classical background
field.
We address to the following question:
What will be the effective Lagrangian
for the classical background field interacting with the 
vacuum of TWO quantum fields?
This task is achieved for the $\phi F \tilde{K}$
interaction by integrating out the $F$ and $\phi$ fields
one after the other.
The $K$ field is treated as a classical background electromagnetic field.

The evaluation of the effective Lagrangian for
a classical background field
interacting with the vacuum of a quantum field 
originated in the works of
W. Heisenberg and H. Euler\cite{heisen} and V. Weisskopf\cite{weiss}.
An explicit expression for the effective Lagrangian  for
the electromagnetic field interacting with the vacuum of the fermion
field (by integrating out the fermion field
in the QED Lagrangian) was evaluated by Julian Schwinger 
in \cite{sch1951}.
Schwinger introduced the proper time method for this purpose
which involved solving for the `dynamics' of a `particle'
with its space-time coordinates evolving with respect to a proper time 
variable.
Schwinger's result has been reproduced using other techniques by various 
authors\cite{itzyk,naroz,popov,duff,neuberger,nussinov,casher,strayer,wang,balantekin,paddy}. 
Systematic analysis of the formalisms and references to related 
works can be found in \cite{fradkin,dittrich}.
Schwinger mechanism has attracted considerable attention because
of its wide range of applicability.
Recently in the SLAC E144 experiment electron positron pairs by
Schwinger mechanism are reported to have been produced
in the collision of intense laser beams with a high energy
electron beam\cite{burke}.
Schwinger mechanism has been used to study particle creation
by an external gravitational field to understand their implications
on Cosmological models \cite{davies}.
There is a strong conviction that
the Hawking's theory of particle creation by black holes
\cite{hawk} can be understood using the Schwinger mechanism.
In \cite{grifols} Schwinger mechanism has been used to put bounds
on the masses of particles.

Schwinger's result 
was derived from the QED Lagrangian for the case of
a constant classical  electromagnetic field.
Various generalizations and extensions to the Schwinger's
result have been attempted by different authors.
Generalization of the Schwinger's result for various other 
interactions involving two fields was done in \cite{duff}.
Generalization for non abelian fields in the context of 
the QCD vacuum breaking down into quark antiquark pairs in the
presence of background colour fields was studied in \cite{cox}.
Schwinger mechanism has been extended to include finite temperature
effects in \cite{rojas}.
Extension for the case where the background electromagnetic 
field is confined to a finite volume has been studied in 
\cite{wang}.

In this work we evaluate the effective Lagrangian for a
classical background electromagnetic field $K$ interacting
with the vacuum of the quantized pseudoscalar field $\phi$
and the quantized electromagnetic field $F$.
The interaction term reads as $\phi F \tilde{K}$.
The evaluation of the effective Lagrangian involves the integration
of the $\phi$ and $F$ fields one after the other.
The effective Lagrangian for the $\phi F \tilde{K}$ interaction
has been earlier evaluated in \cite{grifols}, where the emphasis
has been to use the result to put a bound on the mass of the
pseudoscalar particle called axion. The calculation in 
\cite{grifols} for the evaluation of the effective Lagrangian
is not completely satisfactory because the integration of the $F$ field
in the calculation is qualitatively evaluated using
Feynman diagram techniques. The main goal 
of this paper is to complete this gap by giving the complete
calculation for the integration of the $F$ field.
This part of the calculation
is very important if one intends to generalize the result 
for other interactions. 
 
In section \ref{formalism} we formulate the problem and 
introduce the formalism used.
In section \ref{Aint} we evaluate the $F$ integral.
We shall explicitly show that the contribution to the effective
Lagrangian obtained after the integration of the $F$ field is not non-local
because one of the integrals leads to a delta function.
In section \ref{phiint} we evaluate the 
integration involving the $\phi$ field 
to arrive at the expression for the effective Lagrangian 
under consideration.
Here we use the Schwinger's proper time method to evaluate
the integration of the $\phi$ field.
We show that the result obtained by using the proper time method 
exactly matches with the result obtained in \cite{grifols} using
the Green's function method which was more in the spirit of 
Brown and Duff's paper\cite{duff}.
The complete calculation of the effective Lagrangian can be summarized as:

\begin{minipage}{5in}
\begin{center}
\bea
{\cal {L}} [\phi , F, K]
&=&
- \frac{1}{4} K^2
- \frac{1}{4} F^2
+ \frac{1}{2} (\partial \phi)^2 
- \frac{1}{2} m^2 \phi^2
+ \frac{1}{2} g \phi F \tilde{K}
\nonumber
\eea
\bea
\hspace{20mm}
&&
\left\downarrow
\begin{array}{c} ~ \\ ~ \end{array}
F ~~\mbox{integration}
\right.
\nonumber
\eea
\bea
{\cal {L}}_{eff}' [\phi , K]
&=&
- \frac{1}{4} K^2
+ \frac{1}{2} (\partial \phi)^2
- \frac{1}{2} m^2 \phi^2
- \frac{1}{8} g^2 K^2 \phi^2
\nonumber
\eea
\bea
\hspace{20mm}
&&
\left\downarrow
\begin{array}{c} ~ \\ ~ \end{array}
\right.
\phi ~~\mbox{integration}
\nonumber
\eea
\bea
{\cal {L}}_{eff} [K]
&=&
- \frac{1}{4} K^2
+
\frac{1}{32 \pi^2}
\int_0^{\infty}
\frac{ds}{s^3}
e^{- m^2 s}
        \left[
        \left\{
          \mbox{det}
          \left(
            \frac{g K_1 s}
                 {\sin g K_1 s}
          \right)
        \right\}^{\frac{1}{2}}
        e^{
                - x^2 g K_1 \tan \left( \frac{g K_1 s}{2} \right)
          }
        - 1
        \right]
\nonumber
\eea
\end{center}
\end{minipage}

\noindent
where $K_1$ is the gradient of the electromagnetic field.
As a corollary, in section \ref{pair} we find the expression for 
the probability of the vacuum to breakdown into 
pseudoscalar photon pairs in the presence of a
classical background electromagnetic field.
This is related to the imaginary part of the effective 
Lagrangian. We conclude that the gradient of the 
electric field beyond a certain threshold can give rise to
pair production. This is unlike the QED case where a constant
electric field contributes to pair production.

%------------------------------------------------------------

\section{Formulation of the problem}
{\label{formalism}}

The Lagrangian for a pseudoscalar field $\phi (x)$ interacting 
with a massless vector field $A^{\prime}_{\mu} (x)$ is given by
\bea
{\cal {L}} [\phi , A^{\prime}]
&=&
- \frac{1}{4} {F^{\prime}}_{\mu \nu} {F^{\prime}}^{\mu \nu}
+ \frac{1}{2} ( \partial_{\mu} \phi ) ( \partial^{\mu} \phi)
- \frac{1}{2} m^2 \phi^2
+ \frac{1}{4} g \phi {F^{\prime}}_{\mu \nu}
  \tilde{{F^{\prime}}}^{\mu \nu}
.
\label{orilag}
\eea
The factor of $\frac{1}{2}$ in the kinetic term of the 
pseudoscalar field is to take care of the fact that
we are not treating $\phi$ as a complex field.
In eqn. (\ref{orilag})
${F^{\prime}}_{\mu \nu} (x)$ is the field tensor corresponding to
${A^{\prime}}_{\mu} (x)$ and $\tilde{{F^{\prime}}}^{\mu \nu} (x)$
is the dual tensor corresponding to ${F^{\prime}}_{\mu \nu} (x)$
given by
\be
\tilde{{F^{\prime}}}^{\mu \nu}
= \frac{1}{2} \varepsilon^{\mu \nu \alpha \beta}
{F^{\prime}}_{\alpha \beta}
.
\ee

We shall now show that under the circumstance 
when the $A^{\prime}$ field can be treated as a superposition of
two fields the $\phi F^{\prime} \tilde{F^{\prime}}$ interaction can be
approximated to $\phi F \tilde{K}$ interaction. 
Let us consider the massless vector field $A^{\prime} (x)$
to be a superposition of two independent massless vector fields
given by
\be
{A^{\prime}}^{\mu} (x) \equiv A^{\mu} (x) + a^{\mu} (x)
,
\ee
which amounts to
\bea
{F^{\prime}}^{\mu \nu} (x)
&=& F^{\mu \nu} (x) + K^{\mu \nu} (x)
,
\eea
where $F^{\mu \nu} (x)$ is the field tensor corresponding to the 
field $A^{\mu} (x)$, and $K^{\mu \nu} (x)$ 
is the field tensor corresponding 
to the field $a^{\mu} (x)$.
Let the energy corresponding to the $a^{\mu} (x)$ field 
be $E_a$. Because of the interaction it has with the 
$A^{\mu} (x)$ and $\phi (x)$ field the energy of the 
$a^{\mu} (x)$ field gets modified to $E_a + E_{back}$.
Let us {\it assume} that the correction $E_{back}$ is negligible
compared to $E_a$. That is we make the assumption
\be
E_{back} ~<<~ E_a.
\label{approx}
\ee
In the above approximation we are neglecting the backreaction
of the $A^{\mu} (x)$ and $\phi (x)$ field on the $a^{\mu} (x)$ field.
This is analogous to the case in quantum electrodynamics when
we neglect the radiation effects (back reaction of the electron
on the electromagnetic field).
Next we notice that
neglecting backreaction effects (radiation effects) amounts to
treating one of the fields as a classical field because 
radiation effects becomes significant only in the quantum 
regime (for distances below $\frac{\hbar}{m c}$, where $m$ is
the mass of the particle under consideration).
That is to say, if we are neglecting the backreaction effects
we can as well treat the field to be a classical field. 
Thus we shall interpret $a^{\mu} (x)$ as 
a classical background electromagnetic field.
$A^{\mu} (x)$ and $\phi (x)$ will be treated as quantum fields.

Including the above considerations in
the interaction term of the Lagrangian in eqn. (\ref{orilag})
we have
\bea
{\cal {L}}_{int} [\phi , A , a ]
&=&
\frac{1}{4} g \phi
( F_{\mu \nu} + K_{\mu \nu} )
( {\tilde{F}}^{\mu \nu} + {\tilde{K}}^{\mu \nu} )
\nonumber \\
&=&
\frac{1}{4} g \phi F_{\mu \nu} {\tilde{F}}^{\mu \nu}
+ \frac{1}{2} g \phi F_{\mu \nu} {\tilde{K}}^{\mu \nu}
+ \frac{1}{4} g \phi  K_{\mu \nu} {\tilde{K}}^{\mu \nu}
,
\eea
where we have used the relation
$K_{\mu \nu} \tilde{F}^{\mu \nu}
= F_{\mu \nu} \tilde{K}^{\mu \nu}$.
Since we shall be interested in the interaction of 
all the three fields together, i.e. the 
pseudoscalar field and the quantum electromagnetic field 
in the presence of a classical background electromagnetic field,
we shall {\it ignore} the 
$\phi F \tilde{F}$ and the 
$\phi K \tilde{K}$ terms.
This is done to simplify the calculations.
Thus we have the Lagrangian to be
\bea
{\cal {L}} [\phi , A, a]
&=&
- \frac{1}{4} F_{\mu \nu} F^{\mu \nu}
+ \frac{1}{2} ( \partial_{\mu} \phi ) ( \partial^{\mu} \phi)
- \frac{1}{2} m^2 \phi^2
+ \frac{1}{2} g \phi F_{\mu \nu} \tilde{K}^{\mu \nu}
,
\eea 
where $\tilde{K}$ involves the 
classical background electromagnetic field $a_{\mu} (x)$. 
The corresponding action is given by
\bea
S[\phi , A , a ]
&=&
\int d^4 x ~
{\cal L} [\phi , A , a ]
.
\eea
The vacuum to vacuum transition amplitude is given by
\bea
W
&=&
\frac{
\int {\cal D} {\phi} \int {\cal D} A~ 
        e^{iS [\phi , A , a ]}
}{
\int {\cal D} {\phi} \int {\cal D} A~ 
        e^{iS_0 [\phi , A , a ]}
}
\label{eqn8}
\eea
where $S_0 [\phi , A , a ]$ is the Action in the absence 
of any interaction.
We define the effective action corresponding to the approximation
in eqn. (\ref{approx}) as
\bea
W &=& e^{i S_{eff} [a]}
.
\eea
Note that the effective action is a function of the 
$a^{\mu} (x)$ field alone.
Thus evaluation of the effective action will involve
the integration of the $\phi$ field and the $A$ field
in eqn. (\ref{eqn8}).
In section \ref{Aint} we evaluate the integration of the $A$ field and in
section \ref{phiint} we evaluate the integral involving the $\phi$ field.

The probability for the vacuum to breakdown into pairs is given by
\bea
P &=& 1 - \left| W \right|^2
\nonumber \\
&=& 1 - \exp
        \left[
        - 2~\mbox{Im}~ S_{eff} [a]
        \right]
.
\label{eqn9}
\eea
In section \ref{pair} we shall use this expression to calculate 
the probability for the vacuum to breakdown into pseudoscalar
photon pairs in the presence of a classical background
electromagnetic field.
 
%-------------------------------------------------------------

\section{Evaluation of the $A$ integration}
{\label{Aint}}

In this section we shall begin with eqn. (\ref{eqn8}) and
evaluate the integral involving the $A$ field.
The $A$ integral is first converted into a Gaussian integral.
The contribution to the effective Lagrangian obtained after 
the evaluation of the Gaussian integral on the first look
seems to be non-local. We shall explicitly show that this contribution
is not non-local by showing that one of the integrals leads to
a delta function.
In \cite{grifols} this part of the calculation was evaluated
qualitatively using Feynman diagram techniques. 

To begin with we rewrite eqn. (\ref{eqn8}) as
\bea
W
&=&
\frac{
\int {\cal D} {\phi}~
e^{ i \int d^4 x~
\left[
\frac{1}{2} ( \partial_{\mu} \phi ) ( \partial^{\mu} \phi)
- \frac{1}{2} m^2 \phi^2
\right] }
\int {\cal D} A~ 
e^{ i \int d^4 x~
\left[
- \frac{1}{4} F_{\mu \nu} F^{\mu \nu}
+ \frac{1}{2} g \phi F_{\mu \nu} {\tilde{K}}^{\mu \nu}
\right] }
}{
\int {\cal D} {\phi}~
e^{ i \int d^4 x~
\left[
\frac{1}{2} ( \partial_{\mu} \phi ) ( \partial^{\mu} \phi)
- \frac{1}{2} m^2 \phi^2
\right] }
\int {\cal D} A~ 
e^{ i \int d^4 x~
\left[
- \frac{1}{4} F_{\mu \nu} F^{\mu \nu}
\right] }
}
.
\label{eqn2.5}
\eea
After integration by parts ($x$ integral) 
the $A$ integral in the numerator can be rewritten as
\bea
\int {\cal D} A~ 
e^{ i \int d^4 x~
\left[
- \frac{1}{2}
A_{\mu} \left( 
	  \partial^{\mu} \partial^{\nu} 
	  - g^{\mu \nu} \partial_{\alpha} \partial^{\alpha}
 	\right) A_{\nu}
+ \left\{
    g \partial_{\nu} 
    \left( \phi (x) \tilde{K}^{\mu \nu} (x) \right)
  \right\} A_{\mu}
\right] } 
.
\label{eqn2.9}
\eea
The Gaussian integral formula for a massless vector field
is given by
\bea
\int {\cal D} A~
e^{ i \int d^4 x~
\left[
- \frac{1}{4} F_{\mu \nu} (x) F^{\mu \nu} (x)
+ J_{\mu} (x) A^{\mu} (x)
\right]
}
~~~~~~~~~~~~~~~~~~~~~~~~~~~~~~~~~~~~~~~~~~~~~~~~~~~~~~~~~~
\nonumber \\
~~~~~~~~~~~~
= e^{ - \frac{1}{2}
\int d^4 x~ \int d^4 {x^{\prime}}~
J_{\mu} (x) G^{\mu \nu} ( x - {x^{\prime}} ) J_{\nu} ({x^{\prime}})
}
\int {\cal D} A~
e^{ i \int d^4 x~
\left[
- \frac{1}{4} F_{\mu \nu} (x) F^{\mu \nu} (x)
\right]
}
\label{eqn2.10}
\eea
where
\bea
G^{\mu \nu} ( x - {x^{\prime}} )
&=& \frac{i}{(2 \pi)^4}
\int d^4 k~ 
\left[
k^{\mu} k^{\nu} - g^{\mu \nu} k^2
\right]^{-1}
e^{- i k (x - {x^{\prime}})}
\label{eqn2.11}
\eea
Using eqn. (\ref{eqn2.10}) and eqn. (\ref{eqn2.9})
in eqn. (\ref{eqn2.5}) we have
\bea
W
&=&
\frac{
\int {\cal D} {\phi}~
e^{ i \int d^4 x
\left[
\frac{1}{2} ( \partial_{\mu} \phi ) ( \partial^{\mu} \phi )
- \frac{1}{2} m^2 \phi^2
\right]  
- \frac{1}{2} \int d^4 x \int d^4 {x^{\prime}}~
\left[
g \partial_{\alpha} \left( \phi (x) \tilde{K}^{\mu \alpha} (x) \right)
\right]
G_{\mu \nu} ( x - {x^{\prime}} )
\left[
g {\partial^{\prime}}_{\beta}
\left( \phi ({x^{\prime}}) \tilde{K}^{\nu \beta} ({x^{\prime}}) \right)
\right] }
}{
\int {\cal D} {\phi}~
e^{ i \int d^4 x~
\left[
\frac{1}{2} ( \partial_{\mu} \phi ) ( \partial^{\mu} \phi )
- \frac{1}{2} m^2 \phi^2
\right] }
}
.
\label{eqn2.18}
\eea
Again by integration by parts we can show that
the integral in the interaction term in the numerator is 
\bea
\int d^4 x~
\left[
g \partial_{\alpha} \left( \phi (x) \tilde{K}^{\mu \alpha} (x) \right)
\right]
\int d^4 {x^{\prime}}~
G_{\mu \nu} ( x - {x^{\prime}} )
\left[
g {\partial^{\prime}}_{\beta}
\left( \phi ({x^{\prime}}) \tilde{K}^{\nu \beta} ({x^{\prime}}) \right)
\right]
~~~~~~~~~~~~~~~~~~
\nonumber \\
~~~~~~~~~
=
g^2 \int d^4 x~
\phi (x) \tilde{K}^{\mu \alpha} (x)
\int d^4 {x^{\prime}}~
\left[
\partial_{\alpha} {\partial^{\prime}}_{\beta}
G_{\mu \nu} ( x - {x^{\prime}} )
\right]
\tilde{K}^{\nu \beta} ({x^{\prime}}) \phi ({x^{\prime}})
.
\label{eqn2.20}
\eea
Using eqn. (\ref{eqn2.20}) in eqn. (\ref{eqn2.18}) we have
\bea
W &=&
\frac{
\int {\cal D} {\phi}~
e^{ i \int d^4 x
\left[
\frac{1}{2} ( \partial_{\mu} \phi ) ( \partial^{\mu} \phi )
- \frac{1}{2} m^2 \phi^2
- 
\frac{1}{2} g^2
\int d^4 x \int d^4 {x^{\prime}}~
\phi (x) 
M(x, {x^{\prime}})
\phi ({x^{\prime}})
\right] }
}{
\int {\cal D} {\phi}~
e^{ i \int d^4 x~
\left[
\frac{1}{2} ( \partial_{\mu} \phi ) ( \partial^{\mu} \phi )
- \frac{1}{2} m^2 \phi^2
\right]  }
}
\label{eqn2.22}
\eea
where
\be
M(x, {x^{\prime}})
\equiv
\tilde{K}^{\mu \alpha} (x)
\left[
\partial_{\alpha} {\partial^{\prime}}_{\beta}
G_{\mu \nu} ( x - {x^{\prime}} )
\right]
\tilde{K}^{\nu \beta} ({x^{\prime}})
.
\label{eqn2.23}
\ee
Thus the effective Lagrangian obtained after integrating out the 
$A$ field seems to have a non local contribution from 
$M(x, {x^{\prime}})$. 
But we shall show that the $M(x, {x^{\prime}})$
involves a delta function and thus show that the term is
not non local.

We shall now show that
$M(x, {x^{\prime}})$ takes the following form in terms of
the delta function.
\bea
M(x, {x^{\prime}} )
&=&
- \frac{i}{4}
\tilde{K}^{\mu \nu} (x)
\tilde{K}_{\mu \nu} ({x^{\prime}})
\delta^4 (x - {x^{\prime}} )
.
\label{eqn2.24}
\eea
To arrive at the above expression we begin with eqn. (\ref{eqn2.23})
and using eqn. (\ref{eqn2.11}) we have
\bea
M(x, {x^{\prime}} )
&=&
\tilde{K}^{\mu \alpha} (x)
\tilde{K}^{\nu \beta} ({x^{\prime}})
\frac{i}{(2 \pi)^4}
\frac{\partial}{\partial x^{\alpha}}
\frac{\partial}{\partial {x^{\prime}}^{\beta}}
\int_{- \infty}^{+ \infty} d^4 k~
\left[
k_{\mu} k_{\nu} - g_{\mu \nu} k^2
\right]^{-1}
e^{- i k (x - {x^{\prime}})}
.
\eea
Completing the differentiations with respect to 
$x$ and $x^{\prime}$ and using
\be
\left[
k_{\mu} k_{\nu} - g_{\mu \nu} k^2
\right]^{-1}
=
\frac{1}{k^4}
( k_{\mu} k_{\nu} - g_{\mu \nu} k^2 )
\ee
we have
\bea
M(x, {x^{\prime}})
&=&
i~ \tilde{K}^{\mu \alpha} (x)
\tilde{K}^{\nu \beta} ({x^{\prime}})
\frac{1}{(2 \pi)^4}
\int_{- \infty}^{+ \infty} d^4 k~
( k_{\mu} k_{\nu} - g_{\mu \nu} k^2 )~
\frac{k_{\alpha} k_{\beta}}{k^4}~
e^{- i k (x - {x^{\prime}})}
.
\label{eqn2.26}
\eea
Only the even terms in the $d^4 k$ integral in the above 
expression give a non zero contribution.
Thus we can evaluate the integral by replacing
$k_{\alpha} k_{\beta}$ with 
$\frac{1}{2} g_{\alpha \beta} k^2$.
Using these substitutions in eqn. (\ref{eqn2.26}) we have
\bea
M(x, {x^{\prime}})
&=&
- \frac{i}{4}
\tilde{K}^{\mu \nu} (x)
\tilde{K}_{\mu \nu} ({x^{\prime}})
\frac{1}{(2 \pi)^4}
\int d^4 k~
e^{- i k (x - {x^{\prime}})}
.
\label{eqn2.30}
\eea
Using
\be
\delta^4 (x - {x^{\prime}})
= \frac{1}{(2 \pi)^4}
\int d^4 k~
e^{- i k (x - {x^{\prime}})}
\ee
in eqn. (\ref{eqn2.30}) we have the result
\bea
M(x, {x^{\prime}})
&=&
-\frac{i}{4}
\tilde{K}^{\mu \nu} (x)
\tilde{K}_{\mu \nu} ({x^{\prime}})
\delta^4 (x - {x^{\prime}} )
.
\label{eqn2.31}
\eea

Completing the $x^{\prime}$ integral after substituting
eqn. (\ref{eqn2.31}) in eqn. (\ref{eqn2.22}) and then using
the relation 
$\tilde{K}^{\mu \nu} (x) \tilde{K}_{\mu \nu} (x)
= - K^{\mu \nu} (x) K_{\mu \nu} (x)$ we get
\bea
W
&=&
\frac{
\int {\cal D} {\phi}~
e^{ i \int d^4 x
\left[
\frac{1}{2} ( \partial_{\mu} \phi ) ( \partial^{\mu} \phi )
- \frac{1}{2} m^2 \phi^2
- \frac{1}{8} g^2
\phi (x)
K^{\mu \nu} (x)
K_{\mu \nu} (x)
\phi (x)
\right]
}
}{
\int {\cal D} {\phi}~
e^{ i \int d^4 x~
\left[
\frac{1}{2} ( \partial_{\mu} \phi ) ( \partial^{\mu} \phi )
- \frac{1}{2} m^2 \phi^2
\right]  }
}
.
\label{eqn2.40}
\eea
It should be observed that the expression in the exponential
of eqn. (\ref{eqn2.40}) is the effective Lagrangian for the 
case when both the $\phi$ and $K$ fields are treated as
classical background fields.

%-------------------------------------------------------------

\section{Evaluation of the $\phi$ integral}
{\label{phiint}}

In this section we shall begin with eqn. (\ref{eqn2.40})
and evaluate the integral involving the $\phi$ field.
In \cite{grifols} this was 
achieved by using the Green's function method
adopted by Brown and Duff in \cite{duff}.
Here we shall do the same using the proper time method
introduced by Schwinger in \cite{sch1951}.
Proper time method reduces the problem into an `associated
dynamical problem' with respect to a proper time variable.
The effective Lagrangian in this method is given in terms of the trace of
an evolution operator (evolving in the proper time).
The evolution operator is determined by solving
the `heat equation' corresponding to the `associated
dynamical problem'.
We shall show that the result obtained by using the proper time
method here exactly matches with the result 
obtained in \cite{grifols} using the Green's
function method.

To begin with we rewrite eqn. (\ref{eqn2.40}) as
\bea
e^{i \int d^4 x~ {\cal L}_{eff}(x)}
&=&
\frac{
\int {\cal D} {\phi}~
e^{ i \int d^4 x~
\phi (x)
\left[
- \frac{1}{2} \partial_{\mu} \partial^{\mu}
- \frac{1}{2} m^2 
- \frac{1}{8} g^2 K^2 (x)
\right]
\phi (x)
}
}{
\int {\cal D} {\phi}~
e^{ i \int d^4 x~
\phi (x)
\left[
- \frac{1}{2} \partial_{\mu} \partial^{\mu}
- \frac{1}{2} m^2 
\right] \phi (x)  }
}
\label{eqn3.10}
\eea
where $K^2 = K^{\mu \nu} K_{\mu \nu}$. 
Using the Gaussian integral formula for a scalar field given by
\bea
\int {\cal D} {\phi}~
e^{ i \int d^4 x~ \phi(x) ~\mbox{\boldmath M}~ \phi(x)}
&=&
\mbox{const}
\left[ \mbox{det} ~\mbox{\boldmath M} \right]^{- \frac{1}{2}}
\label{eqn3.11}
\eea
where $M$ is any operator
%Using eqn. (\ref{eqn3.11}) in eqn. (\ref{eqn3.10}) 
and exploiting the identity 
$\mbox{det}~\mbox{\boldmath M} = \exp[\mbox{Tr}~ \ln \mbox{\boldmath M}]$
we have
\bea
\int d^4 x~ {\cal L}_{eff}(x)
&=&
\frac{i}{2}
~\mbox{Tr}
\ln \left\{
	\frac{1}{2} \partial_{\mu} \partial^{\mu}
	+ \frac{1}{2} m^2
	+ \frac{1}{8} g^2 K^2 (x)	
    \right\}
- \frac{i}{2}~\mbox{Tr}
\ln \left\{
        \frac{1}{2} \partial_{\mu} \partial^{\mu}     
        + \frac{1}{2} m^2
    \right\}
.
\label{eqn3.15}
\eea

Thus the evaluation of the effective Lagrangian involves the 
determination of the trace of the logarithm of an operator.
We shall now very briefly give the prescription for the 
determination of the trace of the logarithm of an operator.
Let us begin with the integrals
\bea
\ln y 
= \int_1^y dt~ \frac{1}{t}
\hspace{10mm}
\mbox{and}
\hspace{10mm}
\frac{1}{t}
= \int_0^{\infty} ds~ e^{-t s}
\hspace{10mm}
t > 0
.
\eea
Generalizing these integrals for an operator we have the integral 
representation for $\ln \mbox{\boldmath M}$ to be
\bea
\ln \mbox{\boldmath M}
&=&
- \int_0^{\infty} \frac{ds}{s}
~e^{- \mbox{\boldmath M} s}
.
\eea
Taking the trace of the above integral on both sides and evaluating
the trace in the position basis we have
\bea
\mbox{Tr} \ln \mbox{\boldmath M} 
&=&
- \int dx~
\int_0^{\infty} \frac{ds}{s}~
\left\{
  K(x, x^{\prime}; s)
\right\}_{x = x^{\prime}}
\label{eqn3.17}
\eea
where
$
K(x, x^{\prime}; s)
=
\exp \left[- \mbox{\boldmath M} s \right] \delta (x - x^{\prime}).
%\label{eqn3.20}
$
By differentiating 
%eqn. (\ref{eqn3.20}) 
$
K(x, x^{\prime}; s)
=
\exp \left[- \mbox{\boldmath M} s \right] \delta (x - x^{\prime})
$
with respect to $s$ on both sides we get the differential equation for 
$K(x, x^{\prime}; s)$ to be
\bea
\mbox{\boldmath M} ~ K(x, x^{\prime}; s)
&=&
- \frac{\partial}{\partial s} K(x, x^{\prime}; s)
\label{eqn3.21}
\eea
with the initial condition 
$K(x, x^{\prime}; 0) = \delta (x - x^{\prime})$.
Thus $\mbox{Tr} \ln \mbox{\boldmath M}$ 
is given in terms of the trace of an evolution operator
evolving in the proper time $s$.
The evolution operator is to be determined  by solving 
the differential equation in eqn. (\ref{eqn3.21}).

Using the prescription in eqn. (\ref{eqn3.17}) 
in the expression for the effective Lagrangian
in eqn. (\ref{eqn3.15}) we get
\bea
{\cal L}_{eff} (x)
&=&
\frac{i}{2}
\int_0^{\infty} \frac{ds}{s}
\left\{ U(x, x^{\prime}; s) \right\}_{x = x^{\prime}}
- \frac{i}{2}
\int_0^{\infty} \frac{ds}{s}
\left\{ U_0(x, x^{\prime}; s) \right\}_{x = x^{\prime}}
\label{eqn3.22}
\eea 
where $U(x, x^{\prime}; s)$ and $U_0(x, x^{\prime}; s)$
satisfy the differential  equations
\bea
\left\{
	\frac{1}{2} \partial_{\mu} \partial^{\mu}
        + \frac{1}{2} m^2
        + \frac{1}{8} g^2 K^2 (x)
\right\} U(x, x^{\prime}; s)
&=&
- \frac{\partial}{\partial s} U(x, x^{\prime}; s)
\label{eqn3.24}
\eea
and
\bea
\left\{
        \frac{1}{2} \partial_{\mu} \partial^{\mu}
        + \frac{1}{2} m^2
\right\} U_0(x, x^{\prime}; s)
&=&
- \frac{\partial}{\partial s} U_0(x, x^{\prime}; s)
\label{eqn3.25}
\eea
with the initial conditions
$U(x, x^{\prime}; 0) = \delta^4 (x - x^{\prime})$
and
$U_0(x, x^{\prime}; 0) = \delta^4 (x - x^{\prime})$.
Thus the problem of evaluation of the effective Lagrangian 
reduces to solving the differential equations
in eqn. (\ref{eqn3.24}) and eqn. (\ref{eqn3.25}) with the 
appropriate initial conditions.
Eqn. (\ref{eqn3.24}) cannot be solved for a general $K^2 (x)$.
We thus expand $K_{\mu \nu} (x)$ using Taylor expansion and
take only the leading order terms. Thus we can write 
\bea
K_{\mu \nu} (x)
&=&
\mbox{K}_0
+ \mbox{K}_1 \cdot (\mbox{x} - \mbox{x}_0)
+ ...
\eea 
where $\mbox{K}_0 \equiv K_{\mu \nu} (x_0)$
and $\mbox{K}_1 \equiv \partial_{\alpha} K_{\mu \nu} (x_0)$.
$K_0$ corresponds to the constant part of the electromagnetic 
field and $K_1$ is the 
gradient of the electromagnetic field.
Let us choose $x_0$ to be the stationary point of 
$K^2 (x)$ given by
\bea
x_0 &=& \frac{K_0}{K_1}
.
\label{eqn3.28}
\eea
This choice corresponds to the saddle point approximation.
Under this approximation the differential equation 
in eqn. (\ref{eqn3.24}) becomes
\bea
\left\{
        \frac{1}{2} \partial_{\mu} \partial^{\mu}
        + \frac{1}{2} m^2
        + \frac{1}{8} g^2 {K_1^{\mu \nu}}^2 x_{\mu} x_{\nu}
\right\} U(x, x^{\prime}; s)
&=&
- \frac{\partial}{\partial s} U(x, x^{\prime}; s)
.
\label{EQN3.33}
\eea
In appendix \ref{app1} we solve the above differential equation
and get the solution as 
\bea
U(x, x^{\prime}; s)
&=&
- i
\frac{1}{4 \pi^2}
\frac{1}{s^2}
e^{- \frac{m^2}{2} s}
        \left[
          \mbox{det}
          \left(
            \frac{\frac{g K_1 s}{2}}
                 {\sin \frac{g K_1 s}{2}}
          \right)
        \right]^{\frac{1}{2}}
        e^{l(x, x^{\prime}; s)}
\label{eqn3.35}
\eea
where
\bea
l(x, x^{\prime}; s)
&=&
\frac{1}{4}
(x^2 + {x^{\prime}}^2) g K_1 \cot \left( \frac{g K_1 s}{2} \right) 
- \frac{1}{2} x x^{\prime} g K_1 
	\mbox{cosec} \left( \frac{g K_1 s}{2} \right)
.
\eea
The differential equation for $U_0 (x, x^{\prime}; s)$
can also be very easily solved.
The solution is
\bea
U_0(x, x^{\prime}; s)
&=&
- i
\frac{1}{4 \pi^2}
\frac{1}{s^2}
e^{- \frac{m^2}{2} s}
e^{\frac{(x - x^{\prime})^2}{2s}}
.
\label{eqn3.36}
\eea
From the solutions for $U(x, x^{\prime}; s)$
and $U_0(x, x^{\prime}; s)$ it is observed that
\bea
\lim_{K_0 \rightarrow ~0}
~\lim_{K_1 \rightarrow ~0}
U(x, x^{\prime}; s)
&=&
U_0(x, x^{\prime}; s)
\eea
which implies
that in the absence of any interaction the evolution
operator reduces to that of a free field.
Using the solutions for $U(x, x^{\prime}; s)$
and $U_0(x, x^{\prime}; s)$ 
(eqn. (\ref{eqn3.35}) and eqn. (\ref{eqn3.36}))
in the expression for the effective 
Lagrangian in eqn. (\ref{eqn3.22}) we get
\bea
{\cal L}_{eff} (x)
&=&
\frac{1}{32 \pi^2}
\int_0^{\infty}
\frac{ds}{s^3}
e^{- m^2 s}
	\left[
        \left\{
          \mbox{det}
          \left(
            \frac{g K_1 s}
                 {\sin g K_1 s}
          \right)
        \right\}^{\frac{1}{2}}
	\exp \left\{
		- x^2 g K_1 \tan \left( \frac{g K_1 s}{2} \right)
	     \right\}
	- 1
	\right]
\label{eqn3.38}
\eea 
The above expression for the effective Lagrangian
is best applicable for space time points around the 
stationary point 
$x = \frac{K_0}{K_1}$
because of the approximation made in eqn. (\ref{eqn3.28}).

%-------------------------------------------------------------

\section{Pseudoscalar-photon pair creation}
{\label{pair}}

The effective Lagrangian for the 
$\phi F \tilde{K}$ interaction in eqn. (\ref{eqn3.38})
when evaluated around the stationary point
$x = \frac{K_0}{K_1}$  gives
\bea
{\cal L}_{eff}^{\phi F \tilde{K}} (K_0, K_1)
&=&
\frac{1}{32 \pi^2}
\int_0^{\infty}
\frac{ds}{s^3}
e^{- m^2 s}
        \left[
        \left\{
          \mbox{det}
          \left(
            \frac{g K_1 s}
                 {\sin g K_1 s}
          \right)
        \right\}^{\frac{1}{2}}
        e^{ - l(K_0, K_1; s) }
        - 1
        \right]
\label{eqn4.5}
\eea
where
\bea
l(K_0, K_1; s)
&=&
\frac{1}{2} g^2 {K_0}^2 s
~\frac{\tan \left( \frac{g K_1 s}{2} \right)}{\frac{g K_1 s}{2}}
\eea

The probability of pseudoscalar photon pair production is 
related to the effective Lagrangian for the  
$\phi F \tilde{K}$ interaction by the formula
(using eqn. (\ref{eqn9}))
\bea
P^{\phi F \tilde{K}}
&=& 1 - \exp
        \left[
        - 2 V T ~\mbox{Im}~ {\cal L}_{eff}^{\phi F \tilde{K}}
        \right]
\eea
where $V$ is the total volume of the space and $T$ is the total time for
which the system is observed.
Note that $VT$ should be centered about the stationary point
$x = \frac{K_0}{K_1}$.
The expression for the effective Lagrangian for the 
$\phi F \tilde{K}$ interaction in eqn. (\ref{eqn4.5}) involves the 
determinant of $K_1$. Determinant of an operator is equal to the product 
of its eigenvalues. Schwinger in \cite{sch1951} evaluated the 
eigenvalues of the electromagnetic field tensor $K_{\mu \nu}$
using a very simple and elegant trick.
Using the basic relations satisfied by $K_{\mu \nu}$
he constructed an eigenvalue equation satisfied by the 
eigenvalues of $K_{\mu \nu}$ to be
\bea
\lambda^4
+ \left( \mbox{\bf B}^2 - \mbox{\bf E}^2 \right) \lambda^2
- \left( \mbox{\bf B} \cdot \mbox{\bf E} \right)^2
&=&
0
\eea
The roots of the above equation which are the eigenvalues of
$K_{\mu \nu}$ are 
$+ \lambda_r$, $- \lambda_r$, $+ i \lambda_i$, $- i \lambda_i$,
where
\bea
\lambda_r = \sqrt{R} ~\cos \frac{\theta}{2}
\hspace{7mm} &\mbox{and}& \hspace{7mm}
\lambda_i = \sqrt{R} ~\sin \frac{\theta}{2}
.
\eea
$R$ and $\theta$ are constructed as
\bea
R e^{i \theta} 
&\equiv&
\left( \mbox{\bf E}^2 - \mbox{\bf B}^2 \right)
+ i 2 \left( \mbox{\bf E} \cdot \mbox{\bf B} \right)
.
\eea
Thus we have
\bea
R^2
=
\left( \mbox{\bf E}^2 - \mbox{\bf B}^2 \right)^2
+ 4 \left( \mbox{\bf E} \cdot \mbox{\bf B} \right)^2
\hspace{7mm} &\mbox{and}& \hspace{7mm}
\tan \theta
=
\frac{2 \left( \mbox{\bf E} \cdot \mbox{\bf B} \right)}
	{\mbox{\bf E}^2 - \mbox{\bf B}^2}
. 
\eea
Thus one pair of eigenvalues is always real and the other pair is always 
imaginary. This observation is very crucial because we shall soon
see that only the real pair of eigenvalues contribute
to the imaginary part of the effective Lagrangian.
Note that if the electromagnetic field consists of a pure electric 
field then $\lambda_r = \left| \mbox{\bf E} \right|$
and $i \lambda_i = 0$, and if it consists of a pure magnetic
field then $\lambda_r = 0$ and 
$i \lambda_i = i \left| \mbox{\bf B} \right|$.

Using the above results in eqn. (\ref{eqn4.5}) we have
\bea
\mbox{Im}~ {\cal L}_{eff}^{\phi F \tilde{K}} (\lambda_r, \lambda_i)
&=&
\frac{1}{32 \pi^2} \mbox{Im}~
\int_0^{\infty} 
\frac{ds}{s^3} 
e^{- m^2 s}
        \left[
	\frac{g \lambda^1_r s}{\sin~ g \lambda^1_r s}
	~\frac{g \lambda^1_i s}{\sinh~ g \lambda^1_i s}
        ~e^{- l(\lambda_r, \lambda_i; s)}
        - 1
        \right]
\eea
where
\bea
l(\lambda_r, \lambda_i; s)
&=&
g^2 {\lambda^0_r}^2 s
~\frac{\tan \left( \frac{g \lambda^1_r s}{2} \right)}
	{\frac{g \lambda^1_r s}{2}}
~+~ g^2 {\lambda^0_i}^2 s
~\frac{\tan \left( \frac{g \lambda^1_i s}{2} \right)}
        {\frac{g \lambda^1_i s}{2}}
\eea
where $\lambda^0$ and $\lambda^1$ are the coefficients of the
Taylor expansion series of $\lambda$. That is
\bea
\lambda_r
&=&
\lambda^0_r + \lambda^1_r \cdot (x - x_0)
\nonumber \\
\lambda_i
&=&
\lambda^0_i + \lambda^1_i \cdot (x - x_0)
\eea
$\mbox{Im}~ {\cal L}_{eff}^{\phi F \tilde{K}}$ involves integrating
the imaginary part of the above integral.
This integral is not analytic in $\lambda_r$.
Thus we evaluate $\mbox{Im}~ {\cal L}_{eff}^{\phi F \tilde{K}}$
by the prescription
\bea
\mbox{Im}~ {\cal L}_{eff}^{\phi F \tilde{K}} (\lambda_r, \lambda_i)
&=&
\lim_{\epsilon \rightarrow 0}
~\mbox{Im}~ {\cal L}_{eff}^{\phi F \tilde{K}} (\lambda_r - i \epsilon, \lambda_i)
.
\label{eqn4.20}
\eea
The evaluation of the $s$ integral is achieved by extending 
$s$ to the complex plane. The integrand has poles at points satisfying
$g \lambda^1_r s = n \pi$ and 
$g \lambda^1_i s = i n \pi$ for integer $n$. 
The prescription in eqn. (\ref{eqn4.20}) is equivalent to shifting
the poles corresponding to $\lambda^1_r$ below the real line.
The contour is chosen to pass along the real axis. Using these we get
\bea
\mbox{Im}~ {\cal L}_{eff}^{\phi F \tilde{K}} (\lambda_r, \lambda_i)
&=&
\frac{1}{32 \pi^4}
g^2 {\lambda^1_r}^2
\sum_{n=0}^{\infty}
\frac{1}{n^2}
~\exp \left( - \frac{n \pi m^2}{g \lambda^1_r} \right)
~\frac{n \pi \frac{\lambda^1_i}{\lambda^1_r}}
	{\sinh n \pi \frac{\lambda^1_i}{\lambda^1_r} }
~e^{- i l(\lambda_r, \lambda_i)}
\eea
where
\bea
l(\lambda_r, \lambda_i)
&=&
n \pi g ~\frac{{\lambda^0_r}^2}{\lambda^1_r}
~\frac{\tanh \left( \frac{n \pi}{2} \right)}
	{\frac{n \pi}{2}}
~+~
n \pi g ~\frac{{\lambda^0_i}^2}{\lambda^1_r}
~\frac{\tanh \left( \frac{n \pi}{2} \frac{\lambda^1_i}{\lambda^1_r} \right)}
	{\frac{n \pi}{2} \frac{\lambda^1_i}{\lambda^1_r}}
.
\eea
The above expression for the imaginary part of the effective
Lagrangian for the $\phi F \tilde{K}$ interaction in conjunction
with eqn. (\ref{eqn9}) gives the expression for the probability
for the vacuum to breakdown into pseudoscalar photon pairs
in the presence of a classical background electromagnetic field.
The corresponding expression for the $\phi^2 A^2$ interaction 
calculated by Schwinger in \cite{sch1951} in the above notation
will be
\bea
\mbox{Im}~ {\cal L}_{eff}^{\phi^2 A^2} (\lambda_r, \lambda_i)
&=&
\frac{1}{32 \pi^4}
e^2 {\lambda^0_r}^2
\sum_{n=0}^{\infty}
\frac{1}{n^2}
~\exp \left( - \frac{n \pi m_e^2}{e \lambda^0_r} \right)
~\frac{n \pi \frac{\lambda^0_i}{\lambda^0_r}}
        {\sinh n \pi \frac{\lambda^0_i}{\lambda^0_r} }
\eea
where $m_e$ is the mass of the electron and 
$e$ is the charge of the electron. 
The difference in the factor of $2$ with the Schwinger's result 
in \cite{sch1951} is
because we do not consider our scalar field to be a complex field.

Observe that for the $\phi^2 A^2$ interaction 
$\lambda^0_r$ (constant electric field)
is responsible for pair creation.
$\lambda^0_r$ (constant electric field) does not contribute to 
pair creation in the case of $\phi F \tilde{K}$ interaction.
For the $\phi F \tilde{K}$ interaction
$\lambda^1_r$ (gradient of electric field) is
necessary for pair creation.
Thus we conclude that for the $\phi F \tilde{K}$ interaction we
need a varying electric field (in space or time) for pair creation.

%-------------------------------------------------------------

\section{Acknowledgments}

I thank Subhendra Mohanty for introducing the problem to me
and for all the fruitful discussions.
I am grateful to Sudhir K. Vempati  and Rishikesh Vaidya for 
carefully going through the manuscript.

%-------------------------------------------------------------
\appendix
%-------------------------------------------------------------

\section{Solution to the differential equation in eqn. (\ref{eqn3.33})}
{\label{app1}}

Here we solve the differential equation
\bea
\left\{
\frac{1}{2} \partial_{\mu} \partial^{\mu}
+ \frac{1}{2} m^2 
+ \frac{1}{2} \lambda_{\mu \nu}^2 x^{\mu} x^{\nu}
\right\}
U(x, x^{\prime}; s)
&=&
- \frac{\partial}{\partial s} U(x, x^{\prime}; s)
\label{eqnA1.1}
\eea
with the initial condition
$
U(x, x^{\prime}; 0)
=
\delta^4 (x - x^{\prime})
$.
We begin by demanding (guided by the form of the solution
for $U_0 (x, x^{\prime}; s)$) the solution to be 
of the form
\bea
U(x, x^{\prime}; s)
&=&
\exp \left\{
	- x_{\mu} A^{\mu \nu} (s) x_{\nu}
	- B_{\mu} (s) x^{\mu}
	- C(s)
     \right\}
.
\label{eqnA1.5}
\eea
Substituting eqn. (\ref{eqnA1.5}) in eqn. (\ref{eqnA1.1}) 
we get the equations for the unknowns
$\mbox{\boldmath A}(s)$, 
$\mbox{\boldmath B}(s)$
and $\mbox{C}(s)$ to be 
\bea
2 \frac{\partial \mbox{\boldmath A}}{\partial s}
&=& 4 \mbox{\boldmath A}^2 
        + {{\boldmath \lambda}}^2
\nonumber \\
2 \frac{\partial \mbox{\boldmath B}}{\partial s}
&=& 4 \mbox{\boldmath A} \cdot \mbox{\boldmath B}
\nonumber \\
2 \frac{\partial \mbox{C}}{\partial s}
&=& \mbox{\boldmath B} 
                \cdot \mbox{\boldmath B}
        - 2 \mbox{Tr} (\mbox{\boldmath A})
	+ m^2
.
\eea
The solutions to the above equations are
\bea
\mbox{\boldmath A} (s) 
&=&
- \frac{1}{2} {{\boldmath \lambda}} 
        \cot (2 {{\boldmath \lambda}} s + \mbox{C}_1)
\nonumber \\
\mbox{\boldmath B} (s)
&=&
\mbox{C}_2 \mbox{cosec} (2 {{\boldmath \lambda}} s + \mbox{C}_1)
\nonumber \\
\mbox{C} (s)
&=&
- \frac{1}{2} \frac{\mbox{C}_2^2}{{\boldmath \lambda}}
	\cot (2 {{\boldmath \lambda}} s + \mbox{C}_1)
- \frac{1}{2}\, \mbox{Tr}
        \left[ \ln \mbox{cosec} 
	  (2 {{\boldmath \lambda}} s + \mbox{C}_1)
 	\right]
+ \frac{m^2}{2} s
- \frac{1}{2} \ln \mbox{C}_3
\label{eqnA1.10}
\eea
where ${{\boldmath C_1}}$, ${{\boldmath C_2}}$
and $C_3$ are constants with respect to $s$.
Using eqn. (\ref{eqnA1.10}) in eqn. (\ref{eqnA1.5}) we have
\bea
U(x, x^{\prime}; s)
&=&
\frac{1}{4 \pi^2}
e^{- \frac{m^2}{2} s}
	\left[
	  \frac{(2 \pi)^4 \mbox{C}_3}
		{\mbox{det} \left( \sin
	  (2 {{\boldmath \lambda}} s + \mbox{C}_1) 
	\right) }
        \right]^{\frac{1}{2}}
        e^{l(x, x^{\prime}; s)}
\label{eqnA1.25}
\eea
where 
\bea
l(x, x^{\prime}; s)
&=&
\frac{1}{2}
{{\boldmath x^2}}
{{\boldmath \lambda}}
\cot (2 {{\boldmath \lambda}} s + \mbox{C}_1)
- {{\boldmath x}} \mbox{C}_2
\mbox{cosec} (2 {{\boldmath \lambda}} s + \mbox{C}_1)
+ \frac{1}{2}
\frac{\mbox{C}_2^2}{{\boldmath \lambda}}
\cot (2 {{\boldmath \lambda}} s + \mbox{C}_1)
.
\eea
Using the initial condition
$
U(x, x^{\prime}; 0)
=
\delta^4 (x - x^{\prime})
$
and the identity
\bea
\delta (x - x^{\prime})
&=&
\lim_{\sigma \rightarrow 0}
\frac{1}{\sqrt{2 \pi \sigma}}
e^{- \frac{(x - x^{\prime})^2}{2 \sigma}}
\eea
we can fix ${{\boldmath C_1}}$, ${{\boldmath C_2}}$ and $C_3$
to be
\bea
\mbox{C}_1 = 0
\hspace{3mm}
,
\hspace{10mm}
\mbox{C}_2
  = {{\boldmath \lambda}} 
  \cdot
  {{\boldmath x^{\prime}}}
\hspace{10mm}
\mbox{and}
\hspace{10mm}
\mbox{C}_3
  = \frac{1}{(2 \pi)^4} \mbox{det} ({{\boldmath \lambda}})
.
\label{eqnA1.30}
\eea
Using eqn. (\ref{eqnA1.30}) in eqn. (\ref{eqnA1.25})
we have
\bea
U(x, x^{\prime}; s)
&=&
- i
\frac{1}{4 \pi^2 s^2}
e^{- \frac{m^2}{2} s}
	\left[
	  \mbox{det}
	  \left(
	    \frac{{{\boldmath \lambda}} s}
	         {\sin {{\boldmath \lambda}} s}
	  \right)
        \right]^{\frac{1}{2}}
        e^{l(x, x^{\prime}; s)}
\label{eqnA1.35}
\eea
where 
\bea
l(x, x^{\prime}; s)
&=&
\frac{1}{2}
({{\boldmath x}}^2 + {{\boldmath x^{\prime}}}^2)
{{\boldmath \lambda}}
\cot ({{\boldmath \lambda}} s)
- {{\boldmath x}}
{{\boldmath x^{\prime}}}
{{\boldmath \lambda}}
\mbox{cosec} ({{\boldmath \lambda}} s)
.
\eea

%-------------------------------------------------------------

%-------------------------------------------------------------

%\end{multicols}{2}
\end{document}